\documentclass[useAMS,usenatbib,usegraphicx]{mn2e}

\usepackage{ref}
\usepackage{amssymb}

\newcommand{\etal}{{\em et al.}}
\newcommand{\atrous}{{\em \`a trous~}}

\title[Intragroup light in compact groups II.]{Intragroup diffuse light in compact groups of galaxies II. HCG 15, 35 and 51}

\author[Da Rocha \etal]{C. Da Rocha$^{1,2,3}$
\thanks{E-mail: cdarocha@eso.org},
B. L. Ziegler$^{1,2}$\thanks {E-mail: bziegler@eso.org}
and
C. Mendes de Oliveira$^{4}$\thanks {E-mail: oliveira@astro.iag.usp.br}\\
$^{1}$European Southern Observatory, Karl-Schwarzschild-Str. 2, 85748 
  Garching, Germany\\
$^{2}$Institut f\"ur Astrophysik, Friedrich-Hund-Platz 1, 37077
  G\"ottingen, Germany\\
$^{3}$Divis\~ao de Astrof\'{\i}sica, Instituto Nacional de Pesquisas
  Espaciais (INPE/MCT) \\ Av. dos Astronautas 1758, 12227--010,
  S\~ao Jos\'e dos Campos -- SP, Brazil\\
$^{4}$Instituto de Astronomia, Geof\'{\i}sica e Ci\^encias
  Atmosf\'ericas, Universidade de S\~ao Paulo, \\ Rua do Mat\~ao 1226,
  Cidade Universit\'aria, 05508--900, S\~ao Paulo -- SP, Brazil}
\begin{document}

\date{}

\pagerange{\pageref{firstpage}--\pageref{lastpage}} \pubyear{2007}

\maketitle

\label{firstpage}

\begin{abstract}
This continuing study of intragroup light in compact groups of
galaxies aims to establish new constraints to models of formation
and evolution of galaxy groups, specially of compact groups, which
are a key part in the evolution of larger structures, such as
clusters.  In this paper we present three additional groups (HCG
15, 35 and 51) using deep wide field $B$ and $R$ band images observed
with the LAICA camera at the 3.5m telescope at the Calar Alto
observatory (CAHA). This instrument provides us with very stable
flatfielding, a mandatory condition for reliably measuring intragroup
diffuse light.  The images were analyzed with the OV\_WAV package,
a wavelet technique that allows us to uncover the intragroup component
in an unprecedented way. We have detected that 19, 15 and 26\% of
the total light of HCG 15, 35 and 51, respectively, is in the diffuse
component, with colours that are compatible with old stellar
populations and with mean surface brightness that can be as low as
$28.4~{\rm B~mag~arcsec^{-2}}$. Dynamical masses, crossing times
and mass to light ratios were recalculated using the new group
parameters. Also tidal features were analyzed using the wavelet
technique.
\end{abstract}


\section{Introduction}

Matter can be stripped from galaxies during interaction episodes
and form tidal debris (tails, shells and bridges), which, in a dense
environment such as a cluster or a group, will be partially re-absorbed
by the individual systems and partially dispersed by the cluster
tides combined with ram pressure stripping from the hot intracluster
medium on a quick timescale. This dispersed matter will settle into
the cluster potential, forming a very faint component of intracluster
diffuse light (ICL) \citep{mih04}.

This faint component was first observed by \citet{zwi51} in the
Coma cluster and has been observed in several other clusters and
groups since then and also studied through simulations and theoretical
work \citep[see][~for a recent literature review]{vil99,dar05}.
This component can be used to study the dynamical evolution of
structures in a direct way, since it is related to past galaxy
encounters and history of accretion onto the system \citep{dre84},
it then works as an evolutionary clock and it is sensitive to the
dark matter distribution, that has a direct effect on the amount
of stripped matter. However, the correlations between the properties
of the intracluster light and the cluster properties are still not
well defined given the large uncertainties and the few studies in
this area.

More recent simulations \citep*{rud06} show the evolution of the
ICL with time. A large fraction of the observed evolution in their
simulations happens still inside of groups of galaxies that will
subsequently be accreted by the cluster afterwards. This shows that
the preprocessing of the galaxies and of the intergalactic medium,
still in the group environment, strongly influences the properties
of the resulting clusters.  \citet{mur07} also points out that the
group dynamical history, previous to the accretion by the cluster,
is important and may be the reason for the lack of correlation
between the cluster dynamical history and the ICL fraction.  The
simulations of \citet{mur04,mur07} also present a correlation of
the ICL with cluster total mass.  These authors show that the ICL
component has no preferential redshift to start forming, but 70\%
of it is formed after $z \sim 1$ and its stars would have the same
origin as the Brightest Cluster Galaxy's (BCG), at least for $R <
0.5 R_{vir}$.

\citet*{pur07} show that the ``halo'' on cluster scale (where the
ICL would be considered as the cluster halo) originates in massive
satellites ($10^{11}~{\rm M_{\odot}}$) while on galaxy scale the
halo originates in small mass satellites ($10^{8.5}~{\rm M_{\odot}}$),
on the other hand no correlation between the ICL and the total mass
of the cluster was found. \citet*{con07} results agree with
observations, where 80\% of the stripped stars are placed in the
ICL component and the other 20\% settles in the BCG, also showing
the parallel evolution of the ICL and the BCG. According to the the
simulations of \citet{mon06}, the ICL has also implications on the
low evolution of the high-end of the galaxy mass function ($10^{11}~{\rm
M_{\odot}}$) since $z \sim 1$, what is a problem for models of
galaxy formation in $\Lambda$CDM. That can be solved with the
stripping of about 20\% of the total stellar mass.

On the observational side, in an analysis with stacked images of
clusters in the SDSS (Sloan Digital Sky Survey), \citet{zib05} and
\citet{zib07} found that the ICL tends to align with the BCG, so
that radial orbits would be favored. They also found that clusters
with ``faint-BCGs'' ($M_{r,0} > -22.85$), present a suppression of
the ICL that has no correlation with cluster richness.  \citet{cov06}
find that the ICL and the BCG are aligned in Abell 2667 and also
the X-Ray component presents the same behaviour. The deep photometric
study of \citet*{sei07} shows that the outer part of the BCG surface
brightness profile, which would be associated to the ICL, is better
described by an Exponential profile, rather than the usual de
Vaucouleurs profile.

Recent studies \citep{kri07} show an anti-correlation between ICL
fraction and total cluster mass \citep*[also shown by ][]{gon07},
opposite prediction to the expectation by simulations and an expected
correlation, even though weak, between the ICL fraction and the
redshift.

On group scale, simulations of isolated groups \citep{som06} show
that the intragroup light component (IGL) increases with time showing
its ``dynamical clock nature'' and that these structures evolve to
the so called ``fossil groups'' \citep{pon94}.  Fossil groups are
groups in which the magnitude difference between the first and the
second-ranked galaxies is larger than 2 magnitudes and the X-Ray
halo is extended and has a luminosity higher than
$10^{42}~h^{-2}_{50}~{\rm erg~s^{-1}}$ \citep{jon03}.  By construction
of the sample, where all the galaxies should be in a range of three
magnitudes, this kind of structure is avoided in the Hickson Compact
Groups catalogue \citep{hic82}.

The more recent results from observations and simulations are showing
that the evolution of groups of galaxies have significant influence
on the evolution of clusters of galaxies.  Therefore, understanding
the evolution of groups holds the key to explain several effects
we are trying to understand in the present time.  Our focus on this
topic is the study of compact groups of galaxies, and, as in
\citet{dar05}, we bring additional data with the determination of
the IGL in three new groups.

In our previous work, we studied three compact groups from the
Hickson catalogue \citep{hic82}. We have found that the fraction
of the total $B$ band (and $R$ band) light in the IGL component is
$46\pm8$\% ($33\pm6$\%) for HCG 79, $11\pm6$\% ($12\pm2$\%) for HCG
95 and there was no IGL detection for HCG 88. We quote here the new
error estimates for the IGL light fraction for these groups. The
new error estimates are based on new simulations of detection of
IGL with the OV\_WAV, that cover a larger range in signal to noise
ratio. These groups were separated in categories (initial, intermediary
and advances stage of evolution) according to their stage of dynamical
evolution.

Compact groups of galaxies are formed by three to seven galaxies
separated by distances of the order of the galactic diameter (high
projected densities) and with a low velocity dispersion \citep[$\sim
200~{\rm km~s^{-1}}$,][]{hic92}. In this environment the stripping
of material from the galaxies due to interactions, that may give
origin to the IGL component, should be a frequent and efficient
process. The previous studies of this component in compact groups
using photographic plates were only partially successful
\citep*{ros79,pil95a,sul83,mol98} and only using CCDs a more accurate
detection was possible \citep{nis00,whi03}.

Section~\ref{sample} presents a description of the studied groups
HCG 15, 35 and 51. In \S~\ref{data} the observational data and the
reduction procedures applied to the data are presented. In
section~\ref{analysis} we present the analysis of the data and
results. A discussion and summary are presented in \S~\ref{discussion},
where we put together the results from the last paper and this one
to discuss the evolution of groups. Throughout this work we use
$H_0 = 70~{\rm km~s^{-1}~Mpc^{-1}}$, $\Omega_M=0.3$,
$\Omega_{\Lambda}=0.7$.

\section{The Sample} \label{sample}

In this work we studied three more groups from the Hickson Compact
Groups catalogue \citep[HCG --][]{hic82}, selected based on X-ray
detection \citep{pon96} and on angular size. The characteristics
of the studied groups in this work: HCG 15, 35 and 51, are presented
below.

\subsection{HCG 15}

HCG 15 was originally classified as a sextet of galaxies \citep{hic82},
composed of 3 late-type and 3 early-type galaxies.  \citet{hic92}
show that all the 6 galaxies have concordant radial velocities
(inside $\pm~1000~{\rm km~s^{-1}}$ of the median group velocity)
with a mean recession velocity of $6832~{\rm km~s^{-1}}$ ($99.3~{\rm
Mpc}$).  A more recent measurement shows that the elliptical galaxy
HCG 15C has a recession velocity of $9687\pm22~{\rm km~s^{-1}}$
\citep[galaxy UGC 01620 in the Updated Zwicky Catalog - UZC,~][]{fac99},
that puts this galaxy as an interloper in the group.

We will perform the analysis of this group for the two possible
configurations (sextet and quintet, excluding HCG 15C from the
group), in order to test which one is compatible with the stage of
dynamical evolution of the group.

According to \citet{men94} four of the six galaxies of this group
present signs of interaction. The spiral galaxy HCG 15A presents
weak radio emission, the elliptical galaxy HCG 15B shows a dust
lane, HCG 15D, also an elliptical, has non--centric isophotes and
weak radio emission and the Sbc HCG 15F presents a clear disturbed
morphology. This group was detected in X-rays by ROSAT but so far
off-axis that it is not possible to distinguish between intragroup
and individual galaxies emission \citep{mul03} and the total X-ray
luminosity is ${\rm log} L_X = 42.12~{\rm erg~s^{-1}}$ \citep{osm04}.
This group presents an H{\sc i} deficiency of 76\% \citep{ver01}
when compared to the expected amount of H{\sc i}, considering the
member galaxies morphological types and luminosities. The total
H{\sc i} of the groups was estimated based on the sextet configuration.
The scenario where HCG 15C is a distant galaxy projected onto the
group would explain why the galaxy in the center of the group does
not present signs of interaction.

The general properties of HCG 15 are presented in table~\ref{tabdata}.

\subsection{HCG 35}

HCG 35 is also a sextet \citep{hic82} dominated by early-type
galaxies (5 early-type and 1 late-type), that has a mean recession
velocity of $16252~{\rm km~s^{-1}}$ ($241.7~{\rm Mpc}$).

In this group, only HCG 35A was identified with possible signs of
interaction, an S0 galaxy showing weak emission lines that extend
beyond the nucleus \citep{men94}. It has ROSAT X-ray detection but,
as it is also the case of HCG 15, observations do not allow the
distinction between intragroup and individual galaxies emission
\citep{mul03}. Its total X-ray luminosity is ${\rm log} L_X =
42.06~{\rm erg~s^{-1}}$ \citep{pon96}.  This group has an H{\sc i}
overabundance of about 29\% with respect to its total expected H{\sc
i} content \citep{ver01}.

General properties of HCG 35 can be seen in table~\ref{tabdata}.

\subsection{HCG 51}

This group also presents a ``membership issue''. Originally classified
as a sextet of galaxies \citep{hic82}, HCG 51C's radial velocity
is $\sim 1200~{\rm km~s^{-1}}$ above the group's median velocity,
and therefore it was excluded from the group by \citet{hic92}. On
the other hand, as explained by these authors, $\pm 1000~{\rm
km~s^{-1}}$ is an arbitrary threshold, so we decided to analyze
both possibilities.

This group is dominated by an E1 galaxy and has 2 late-type and 4
early-type galaxies and HCG 51C is an S0 galaxy. The mean recession
velocity of this group is $7924~{\rm km~s^{-1}}$ ($112.6~{\rm Mpc}$),
in the quintet configuration.

The only interaction sign identified by \citet{men94} in this group
was a faint disk in the center of the elliptical galaxy HCG 51E,
even though a very low surface brightness jet-like structure
connecting HCG 51A and HCG 51E could be identified in our analysis
(see below) and can be related to the origin of the faint disk. The
total X-ray luminosity of this group is ${\rm log} L_X = 42.70~{\rm
erg~s^{-1}}$ \citep{pon96}. The H{\sc i} deficiency compared to the
expected H{\sc i} content is 69\% \citep{ver01}, estimated for the
quintet configuration.

The general properties of HCG 51, as for the other studied groups
are summarized in table~\ref{tabdata}.

\begin{table}

\centering
\caption{General properties of the observed groups.
\label{tabdata}}

\begin{tabular}{llll}

\hline
           & HCG 15               & HCG 35               & HCG 51 \\
\hline

RA         & ${\rm 02^h07^m39^s0}$&${\rm 08^h45^m19^s5}$&${\rm 11^h22^m20^s9}$\\
DEC        & ${\rm +02^o08'18''}$ &${\rm +44^o31'18''}$ &${\rm +24^o17'35''}$ \\
$V_{Rad}$ (${\rm km~s^{-1}}$)  & 6753/6832* & 16252 & 7728/7924*   \\
Distance (${\rm Mpc}$)         & 98.2/99.3* & 232.2 & 112.6/115.5* \\
$(m-M)_V$                      & 35.0       & 36.9  & 35.3         \\
Vel. Disp. (${\rm km~s^{-1}}$) & 471/462*   & 348   & 265/535*     \\
Mean. Sep. (${\rm kpc}$)       & 118/110*   & 77    & 86/88*       \\
Num. Gal.                & 5/6*       & 6     & 5/6*         \\
\hline
\end{tabular}

{\footnotesize * Values presented for HCG 15 and HCG 51 are given
for the quintet and sextet configurations, respectively.}

{\footnotesize Velocity dispersions were calculated using the
correction for the velocity measurements errors described by
\citet{dan80}.}

\end{table}

\section{Observations and Data reduction} \label{data}

\subsection{Observational data}

The studied groups were observed with the wide--field camera LAICA
(Large Area Imager for Calar Alto) at the prime focus of the 3.5m
telescope of the Calar Alto observatory (CAHA -- Centro Astron\'omico
Hispano Alem\'an).  Deep images were obtained in the $B$ and $R$
bands in December 2003, in non photometric nights, but with
sub-arcsecond seeing (from $0\farcs6$ to $1\farcs0$), with exposure
times ranging between 8250 and 12000 seconds in $B$ and 3500 and
4000 seconds in $R$.  The observational information is summarized
in table~\ref{tabima}.

The camera LAICA is composed of an array of 2$\times$2 CCDs. Each
CCD has $4k\times 4k$ pixels covering a field of view of $15\farcm36
\times 15\farcm36$. CCDs are separated by $13\farcm64$, about the
size of one CCD, so that the usual observational procedure encompasses
4 pointings shifted by one CCD length, resulting in a field of view
of about one degree, covered in 16 separated images. The camera has
a pixel size of $0.225''$, and our observations were performed with
a 2$\times$2 binning, resulting in pixels of $0.45''$.

Since compact groups of galaxies have a small angular size, each
of the observed groups could be fit in one single CCD. This gave
us the good advantage of needing only one single pointing per group,
not the usual four-pointings procedure. While a group is placed in
one single CCD, the other 3 CCDs of the camera were used to observe
blank neighboring regions of the sky, thus producing ``night sky
flats''.  By placing each group in a different CCD we obtained a
complete set of observed science frames with groups and the
corresponding night sky flats with exposure times of the same order.
Since flatfielding is a key issue in the study of large scale low
surface brightness structures, this observational strategy was very
well suited and saved a considerable amount of telescope time.
Individual exposures were dithered to correct for CCD defects and
cosmic rays.

\begin{table}

\centering
\caption{Observational data.
\label{tabima}}

\begin{tabular}{cccc}

\hline
Group & Band   & Exp.  & Seeing \\

      &        & (Sec.)&        \\
\hline

HCG 15 & B & 11250 ($15\times750$)  & $0\farcs8$ \\
       & R & ~4000 ($16\times250$)  & $0\farcs6$ \\
HCG 35 & B & 12000 ($16\times750$)  & $0\farcs9$ \\ 
       & R & ~3500 ($14\times250$)  & $1\farcs0$ \\
HCG 51 & B & ~8250 ($11\times750$)  & $0\farcs9$ \\
       & R & ~4000 ($16\times250$)  & $1\farcs0$ \\

\hline

\end{tabular}
\end{table}

\subsection{Data reduction}

The basic data reduction was performed using IRAF\footnotemark. One
problem encountered was that each CCD of the camera has a four-port
readout, that could affect our analysis that requires stable sky
levels. After performing the BIAS and flatfield corrections the
difference in the level of each quadrant of the CCD did not
disappeared. For this reason we treated each quadrant as a separate
image and made a ``sky level matching'' at the end of the basic
data reduction.

\footnotetext{IRAF (Image Reduction and Analysis Facility) is
distributed by the National Optical Astronomy Observatories, which
is operated by the Association of Universities for Research in
Astronomy, Inc., under cooperative agreement with the National
Science Foundation.}

As mentioned before, due to our observational strategy, we have two
types of images: science images, which are the ``on-group'' ones,
and night sky flats, which are the other three ``off-group'' images,
in each exposure.  Each science and sky flat image was cut in four
quadrants and the BIAS correction was applied to each of the new
``quadrant images''. Due to the non photometric conditions, some
of the ``off-group'' images presented sky patterns and could not
be used to produce the night sky flatfields. Those sky patterns
were sometimes also present in the ``on-group'' images which could
not be corrected by the flatfields and could not be used to produce
our final image. After selecting the useful images, a combination
of night sky flat, to correct for the large scale CCD response and
for CCD illumination, and twilight flat, to correct for the individual
pixel sensitivity was applied to the science frames. The quadrants
were rejoined in a single image with matched sky levels.  The sky
matching consists in measuring the difference in sky level along
both sides of the matching region and correcting the sky level by
a constant, so that no sky level difference is left and no step-like
structure is present in the image.  This correction was at the order
of a few counts, while the sky level is of the order of some thousands
of counts. However, this small variation would have an important
effect in our analysis.  Images were registered and combined producing
a final image for each group in each filter.

Since the nights when the data were taken were not photometric, we
need to calibrate our images by comparing them to published data
by \citet{men92}. But because the \citet{men92} image sizes are
quite small ($2\farcs2~\times~3\farcs5$), we could not measure a
useful number of stars in common for the calibration (the few ones
available in these fields were, in most cases, saturated in our
images).  We, therefore, decided to use the calibrated profiles of
the galaxies themselves, in order to find the zero points for the
images. We matched the instrumental profiles we obtained for each
galaxy with those presented in \citet{men92}, excluding the central
parts, which are usually contaminated by seeing effects, and the
outer parts of the galaxy, which are affected by the sky subtraction.
Even with the excluded regions, several arcseconds of the surface
brightness profiles where used in the fit.  The RMS of the fitting
was quite small (typical RMS = 0.05 magnitudes) but realistic errors
are larger (of the order of 0.1 to 0.2 magnitudes) since they depend
on the reliability of the zero point for the literature data.

\section{Analysis and Results} \label{analysis}

\subsection{Wavelet analysis}

Since the IGL is an extended, very low surface brightness structure
(usually less than 1\% above the night sky level) many instrumental
effects like flatfielding, scattered light and CCD bleeding, among
others, can contaminate the signal. Therefore, very good illumination
corrections and short exposure times to avoid bleeding are necessary.
Besides the instrumental effects, some astronomical ones are important
in this kind of analysis. The dimming of the IGL by the light from
the objects in the image, like member galaxies and stars, which
need to be modeled and subtracted from the image, is one major
``problem'' for this kind of study. Also a very accurate sky
subtraction is needed for the detection and analysis of the IGL.

To deal with this kind of requirement, we applied a wavelet-based
technique, the OV\_WAV package \citep{epi03}, in our analysis, as
shown in \citet{dar05}. This technique does a multiscale analysis
that uses the \atrous wavelet transform, and is able to separate
the structures in the image by their characteristic sizes with no
{\em a priori} information \citep{bij95,sta98}. Since the information
in astronomical images is organized hierarchically (stars projected
onto galaxies, that are projected onto a larger structure like the
IGL and all projected onto the sky brightness level), this technique
is very well suited for our goals. This method was already applied
by other authors to the same kind of study \citep[{\em e.g.}
][]{ada05}.

The main procedure is the following. The images are deconvolved
into wavelet coefficients, which will contain information of a given
size ($2^n$ pixels, where $n$ is the index of the wavelet coefficient
- $n=0, 1, 2, 3, ...$). A source in the image, when deconvolved in
wavelet coefficients, has a representation in each of the coefficients
that depends on the shape and size of this object. Also the noise
present in the image, has a representation in the different wavelet
coefficients. The representation of the noise is analyzed and the
representation of the signal is detected in each coefficient.  These
detections in different coefficients are interconnected, defining
the detected objects and each detected object is reconstructed and
separately analyzed.

The noise present in astronomical images is mostly dominated by
small scale components. As the characteristic size of the wavelet
coefficients increases, the representation of the noise looses
intensity. With this behaviour of the noise representation, working
in wavelet space, we are able to detect large structures, with very
low surface brightness, with high confidence levels.  Our simulations
have shown that we are able to detect at a 5-$\sigma$-detection
level in wavelet space, large, low surface brightness structures,
that have only $S/N = 0.1$ in real space \citep{dar05}.

After the reconstruction of each object, in an iterative process
that models and subtracts the detected sources from the image, we
are able to recompose the IGL component and the galaxies component
for each group, in each band independently, and perform our analysis,
which will be described below. A more detailed description of the
OV\_WAV and the simulations on the detection of IGL can be found
in \citet{dar05}.

\subsection{Group parameters} \label{grpparm}

We have re-calculated the group crossing time ($t_c$) and mass to
light ratio ($M/L$), previously calculated in \citet{hic92}, using
the concordance cosmology parameters, since the original ones were
calculated using $H_0 = 100~{\rm km~s^{-1}~Mpc^{-1}}$ and $\Omega_M=1.0$.

There was very small variations due to cosmology on the re-calculated
values for the crossing time, which is given in unities of ${\rm
H_0^{-1}}$, since for closer groups the impact of $\Lambda$ is
small. On the other hand, including or excluding galaxies from the
analysis, as we did for HCG 15 and HCG 51, can have a considerable
effect on this value. The crossing times were calculated using the
relation presented in \citet{hic92}

\begin{equation}
t_c = \frac{4}{\pi} \frac{R}{D\sigma}.
\end{equation}

R is the mean separation in ${\rm kpc}$ between the member galaxies
and $D\sigma$ is the deprojected velocity dispersion given by

\begin{equation}
D\sigma = [3(<v^2> - <v>^2 - <\sigma_v^2>)]^{1/2},
\end{equation}

\noindent where $v$ is the observed recession velocity and $\sigma_v$
is the measurement error.

For $M/L$, the total group luminosities were calculated with the
member galaxies for each case and the proper distance modulus.
Since X-ray mass estimates are not available for most of groups
from Hickson's catalogue, and for consistent comparison, we used
the same mass estimators of \citet{hic92} to estimate the group's
masses.  The median of the four mass estimators described by
\citet*{hei85} (virial mass, projected mass, median mass and mean
mass) was used for each group. An important consideration should
be raised here, that dynamical mass estimates using the small number
of galaxies present in a compact group can suffer from severe
statistical effects and should be taken only as a guiding value.
These estimators only measure the mass inside the area where the
galaxies are found, what corresponds only to the inner part of the
detected IGL components and of the dark matter halo. Therefore,
these estimators would not account for the mass in the outer parts
of the group.

We have found in the literature new recession velocity measurements
for some of the galaxies, as for the case of HCG 15C, which differ
from the originally published values by about $30~{\rm km~s^{-1}}$,
on average. This small variation causes an impact of up to 20\% in
the estimated group parameters, like crossing times and masses,
showing how sensitive they are to the small number of objects present
in each group. For consistency we decided to use the values from
\citet{hic92}, unless for the case where HCG 15C is not considered
a group member. Using the new recession velocities would not change
the difference between the velocity of HCG 51C and the median
velocity of this group.

The quantities estimated here for each group and its possible
different configurations can be seen in table~\ref{tabparam}.

\begin{table*}

\centering
\caption{Group parameters for each configuration.
\label{tabparam}}

\begin{tabular}{llllll}

\hline
           & HCG 15 (5G) & HCG 15 (6G) & HCG 35 & HCG 51 (5G) & HCG 51 (6G) \\
\hline

$D\sigma$ (${\rm km~s^{-1}}$)  & 726.9 & 730.0 & 547.8 & 409.4 & 844.9 \\
$t_c$ (${\rm H_0^{-1}}$)       & 0.014 & 0.014 & 0.011 & 0.019 & 0.009 \\
M/L (${\rm M_{\odot}/L_{\odot}}$) & 614 & 424 & 55 & 50 & 174 \\
M/L corrected (${\rm M_{\odot}/L_{\odot}}$) & 517 & 367 & 48 & 38 & 139 \\
Group Mass (${\rm M_{\odot}}$) & 5.67$\cdot$10$^{13}$ & 5.30$\cdot$10$^{13}$ & 1.51$\cdot$10$^{13}$ & 7.39$\cdot$10$^{12}$ & 3.59$\cdot$10$^{13}$ \\
E-type fraction          & 0.4 & 0.5 & 0.8 & 0.6 & 0.7 \\

\hline

\end{tabular}

{\footnotesize 5G and 6G designations corresponds to quintet and
sextet configurations of HCG 15 and HCG 51.}

\end{table*}

\subsection{HCG 15}

The images in $B$ and $R$ of HCG 15 were deconvolved into 11 wavelet
coefficients ($2^{10}$ pixels, about the size of the image), the
detected objects were reconstructed in a multiple iterations process
and the group galaxies and IGL component of this group were
re-composed, as is shown in figure~\ref{figdifh15}. The whole
analysis process is made in an totally independent way for the $B$
and $R$ bands.

\begin{figure*}
\centering
\caption{IGL component of HCG 15 identified and reconstructed with
the OV\_WAV package. The left panel shows the image in the $B$ band
and the IGL component as contour curves with surface brightness
levels which range from $27.0$ to $31.0$ magnitudes in steps of
$0.25~{\rm B~mag~arcsec^{-2}}$, from the inner to the outer part.
The right panel shows the image in the $R$ band and the IGL component
as contour curves with surface brightness levels which range from
$25.5$ to $29.0$ magnitudes in steps of $0.25~{\rm R~mag~arcsec^{-2}}$,
from the inner to the outer part.
\label{figdifh15}}
\end{figure*}

For the quintet (and sextet) configuration, the detected IGL component
in this group represents $19\pm4$\% ($16\pm3$\%) of the total light
in the $B$ band and $21\pm4$\% ($18\pm4$\%) in the $R$ band, about
70\% of the light of the first-ranked galaxy.  This component
corresponds to apparent total magnitudes of $B = 15.0\pm0.2$ and
$R = 13.3\pm0.2$, down to a surface brightness detection limit of
$\mu_B = 31.3$ and $\mu_R = 30.5$. The detection limits correspond
to a surface brightness of $0.1\cdot \sigma_{Sky}$ ($S/N = 0.1$)
in each band, which is the detection limit of the OV\_WAV package.

The IGL presents a very irregular shape and has a very low mean
surface brightness, $\mu_B = 28.4\pm0.2$ and $\mu_R = 26.7\pm0.2$.
Its mean $S/N$ per pixel is 1.5 in $B$ and 3.3 in $R$. The IGL has
a mean colour $(B-R)_0 = 1.6\pm0.2$, redder, but still consistent
with the mean colour for the galaxies' component ($(B-R)_0 =
1.4\pm0.2$ for the quintet configuration and $(B-R)_0 = 1.5\pm0.2$
for the sextet) and with the old stellar population expected from
early-type galaxies. The extinction corrections were made using
\citet{rie85} extinction laws and \citet*{sch98} extinction maps.
The IGL and galaxy components were measured in the same areas in
the two different bands. The properties of the IGL components are
summarized in table~\ref{tabres}.

The original $M/L$ in the $B$ band for this group, calculated by
\citet{hic92}, was 432 ${\rm M_{\odot}/L_{\odot}}$. The $M/L$ values
were converted to $H_0 = 70~{\rm km~s^{-1}~Mpc^{-1}}$, while the
original values were calculated with $H_0 = 100~{\rm km~s^{-1}~Mpc^{-1}}$
and $\Omega_M=1.0$.  Our new estimate takes these values to 614 and
424 ${\rm M_{\odot}/L_{\odot}}$, respectively for the quintet and
sextet cases.  Using the light contained in the IGL component, we
recalculated the $M/L$ for each of the studied groups for the new
total luminosity, correcting the estimated values.  In the case of
HCG 15 the $M/L$ drops to 517 and 367 ${\rm M_{\odot}/L_{\odot}}$,
for the two presented cases. The $M/L$ values presented in this
paper always refer to the $B$ band.

The crossing times for this group, 0.014 ${\rm H_0^{-1}}$ \citep{hic92},
had no change in both configurations, quintet and sextet. $M/L$ and
crossing time values are given in table~\ref{tabparam}.

\subsection{HCG 35}

As in the case of HCG 15, the $B$ and $R$ images of HCG 35 were
deconvolved into 11 wavelet coefficients and the galaxies and IGL
component were re-composed after the wavelet multiple iteration
process, as can be seen in figure~\ref{figdifh35}, where the analysis
of the $B$ and $R$ bands are independent.

\begin{figure*}
\centering
\caption{IGL component of HCG 35 identified and reconstructed with
the OV\_WAV package. The left panel shows the image in the $B$ band
and the IGL component as contour curves with surface brightness
levels which range from $27.5$ to and $29.0$ magnitudes in steps
of $0.25~{\rm B~mag~arcsec^{-2}}$, from the inner to the outer part.
The right panel shows the image in the $R$ band and the IGL component
as contour curves with surface brightness levels which range from
$25.75$ to $27.5$ magnitudes in steps of $0.25~{\rm R~mag~arcsec^{-2}}$,
from the inner to the outer part.
\label{figdifh35}}
\end{figure*}

In this group the detected IGL component presents an irregular
shape, elongated in the same direction as the galaxy component
elongation. This component represents $15\pm3$\% of the total light
in the $B$ band and $11\pm2$\% in the $R$ band, 40\% of the light
of the first-ranked galaxy. The total apparent magnitudes of the
IGL are $16.4\pm0.2$ and $14.9\pm0.2$, respectively in the $B$ and
$R$ bands, down to a surface brightness detection limit of $\mu_B
= 31.0$ and $\mu_R = 28.8$.

This group presents quite a faint IGL component with a mean surface
brightness of $\mu_B = 27.9\pm0.2$ and $\mu_R = 26.4\pm0.2$.  The
mean $S/N$ per pixel of the IGL is 1.7 in $B$ and 1.0 in $R$. Its
mean colour is $(B-R)_0 = 1.5\pm0.2$ and the galaxies have a mean
$(B-R)_0 = 1.8\pm0.2$. In this case the IGL component is bluer than
the mean colour for the galaxies, but consistent with the typical
colour for old stellar population in early-type galaxies, while the
galaxy's component is redder than the typical value for early-type
galaxies. The mean colour measured for the galaxies' component is
not surprising considering that, at least 3 galaxies present colours
measured inside the $\mu_R = 24.5$ isophote around $(B-R)_0 = 1.9$,
including one Sb galaxy \citep*{hic89}. IGL properties are shown
in table~\ref{tabres}.

The new estimate of the $M/L$ in the $B$ band takes the value from
57 \citep[][converted to $H_0 = 70~{\rm km~s^{-1}~Mpc^{-1}}$]{hic92}
to 55 ${\rm M_{\odot}/L_{\odot}}$. Correcting it by the light in
the IGL, the $M/L$ drops to 48 ${\rm M_{\odot}/L_{\odot}}$.  $t_c$
slightly increased from 0.010 \citep{hic92} to 0.011 ${\rm H_0^{-1}}$.
The re-estimated parameters for this group can be seen in
table~\ref{tabparam}.

\subsection{HCG 51}

This group's images were analyzed in the same way as the previous
two and the IGL component can be seen in figure~\ref{figdifh51}.

\begin{figure*}
\centering
\caption{IGL component of HCG 51 identified and reconstructed with
the OV\_WAV package. The left panel shows the image in the $B$ band
and the IGL component as contour curves with surface brightness
levels which range from $26.5$ to $29.0$ magnitudes in steps of
$0.25~{\rm B~mag~arcsec^{-2}}$, from the inner to the outer part.
The right panel shows the image in the $R$ band and the IGL component
as contour curves with surface brightness levels which range from
$24.75$ to $27.5$ magnitudes in steps of $0.25~{\rm R~mag~arcsec^{-2}}$,
from the inner to the outer part.
\label{figdifh51}}
\end{figure*}

This group presents a more prominent elliptical shape in the IGL
component, following the elongation of the galaxies distribution
with some clear tidal features. The IGL, in the quintet (and sextet)
configuration, represents $31\pm6$\% ($26\pm5$\%) and $28\pm5$\%
($24\pm5$\%) of the total light in the $B$ and $R$ bands, respectively,
about 70\% of the light of the first-ranked galaxy.  This corresponds
to apparent total magnitudes of $B = 14.4\pm0.2$ and $R = 12.9\pm0.2$,
down to the limiting surface brightnesses of $\mu_B = 30.7$ and
$\mu_R = 29.7$.

The mean surface brightness of this component is $\mu_B = 27.4\pm0.2$
and $\mu_R = 25.9\pm0.2$, The mean $S/N$ is 2.0 in $B$ and 3.4 in
$R$.  The IGL has a mean colour $(B-R)_0 = 1.5\pm0.2$, about the
same as the mean colour of the galaxies, $(B-R)_0 = 1.6\pm0.2$ for
both configurations, as expected for early-type galaxies with old
stellar population. The properties of the IGL component can be seen
in table~\ref{tabres}.

The new estimate of the $M/L$ in the $B$ band was performed for
this group, that goes from 51 \citep[][converted to $H_0 = 70~{\rm
km~s^{-1}~Mpc^{-1}}$]{hic92} to 50 and 174 ${\rm M_{\odot}/L_{\odot}}$,
for quintet and sextet configurations. It drops to 38 and 139 ${\rm
M_{\odot}/L_{\odot}}$, when corrected for the diffuse component
contribution.

The crossing time of this group, originally 0.019 ${\rm H_0^{-1}}$,
presented no change for the quintet configuration and drastically
drops to 0.009 ${\rm H_0^{-1}}$, for the sextet configuration.  The
large difference between the $M/L$ and the crossing time, for the
quintet and sextet configuration, is due to the deprojected velocity
dispersion that is twice as large in the latter case, what can also
be noticed in the 2-D velocity dispersion (tables~\ref{tabdata}
and~\ref{tabparam}).

In HCG 51 we detected a very low surface brightness jet-like structure
that connects galaxies HCG 51A and 51E. This structure is redder
than the mean colour of the galaxy's component and of the IGL
($(B-R)_0 = 1.8\pm0.2$) and has a mean surface brightness $\mu_B =
27.0\pm0.2$ and $\mu_R = 25.2\pm0.2$.  The estimated extension of
this structure is $90~{\rm kpc}$, about the mean galaxy separation,
actually this structure connects the two ``parts'' of this group.
The images also show a possible shell-like structure perpendicular
to the jet-like one, but it is too faint and small to be detected
as an independent object and properly analyzed.  This jet-like
structure can be seen in figure~\ref{figjeth51}. No further tidal
structures were detected in this group.

\begin{figure}
\centering
\caption{Image of HCG 51 in the $B$ band showing the jet-like
structure (marked with the white ellipse) connecting HCG 51A and
51E.
\label{figjeth51}}
\end{figure}

\begin{table*}

\centering
\caption{Properties of the IGL component detected in our sample.
\label{tabres}}
\begin{tabular}{llllllll}

\hline
Group & \multicolumn{2}{c}{\% ($B$ and $R$)} & \multicolumn{2}{c}{$\mu$ ($B$ and $R$)} & \multicolumn{2}{c}{Mag. ($B$ and $R$)} & $(B-R)_0$ \\
\hline

HCG 15  & $19\pm4$\% / $16\pm3$\% & $21\pm4$\% / $18\pm4$\% & $28.4\pm0.2$ & $26.7\pm0.2$ & $15.0\pm0.2$ & $13.3\pm0.2$ & $1.6\pm0.2$ \\
HCG 35  & $15\pm3$\% & $11\pm2$\% & $27.9\pm0.2$ & $26.4\pm0.2$ & $16.4\pm0.2$ & $14.9\pm0.2$ & $1.5\pm0.2$ \\
HCG 51  & $31\pm6$\% / $26\pm5$\% & $28\pm5$\% / $24\pm5$\%  & $27.4\pm0.2$ & $25.9\pm0.2$ & $14.4\pm0.2$ & $12.9\pm0.2$ & $1.5\pm0.2$ \\
\hline
\end{tabular}

Column (1) Group studied.\\
Columns (2-3) Fraction of the group's total light in the IGL component, bands 
$B$ and $R$, respectively \\ 
(the values presented for HCG 15 and 51 are for the quintet and sextet 
configurations) \\
Columns (4-5) Mean surface brightness of the IGL component, bands $B$ and $R$, 
respectively.\\
Columns (6-7) Integrated magnitude of the IGL component, bands $B$ and $R$, 
respectively.\\
Column (8) Extinction corrected colour of the IGL component.
\end{table*}

\subsection{Structural Analysis of the IGL}

As can be seen in figures~\ref{figdifh15} to~\ref{figdifh51}, the
IGL components detected in our study clearly show very irregular
shapes, in contrast with the ICL components detected in clusters
of galaxies \citep[e.g. ][]{kri07}.

In order to quantify how regular, or irregular, the light distribution
can be, we perform the modeling of the IGL component using elliptical
isophotes with the IRAF package {\tt STSDAS.ELLIPSE}.  This analysis
leaves large residuals when the elliptical models are subtracted
from the IGL, even in the case of HCG 51 that shows an apparently
less irregular IGL component.  These large residuals are already
an indication of the lack of regularity of the IGL distribution.

To understand the physical meaning of the fitted elliptical models,
we plotted the surface brightness profiles along the major axis,
which presented a multimodal behavior in all the cases. We tried
to fit four types of light profiles to our data: Exponential, de
Vaucouleurs, King-like core model and Hubble-Reynolds, that should
work as an analytic proxy for a NFW universal profile \citep*{nav97},
as suggested by \citet{lok01} and \citet{kri07}. The profile fitting,
in all cases, either single or double component, led to unphysical
results for parameters as effective radius or scale length (being
extremely small or extremely large), for example, indicating no
dynamical equilibrium.

The unphysical results obtained by the fits, show that the elliptical
models do not represent any physical entity, as expected considering
the large residuals when these models are subtracted. The failure
on fitting the surface brightness profiles is another indication
that the IGL components are very irregular and cannot be described
by a regular model, as it happens in the diffuse component of
clusters of galaxies. We can conclude from this, already, that the
compact groups are far from being a relaxed and virialized structure.

The analysis of the colour maps of the reconstructed IGL components
of all groups do not show any kind of substructures, such as blue
regions that could be associated to star forming regions in the
IGL.  Blue star forming regions in the diffuse component, as found
in the Coma cluster by \citet{ada05}, could be the origin of the
difference we find between the mean colour of the IGL and the one
of the galaxy component, when the IGL is bluer than the galaxies.
In the case of our study, those regions, if they exist, could be
too disperse to be individually identified in our colour maps.

\subsection{Dynamical Evolution Indicators}

Some group properties can be used as dynamical evolution indicators,
such as fraction of early-type galaxies (or of late-type galaxies),
crossing time and magnitude difference between the first and the
second-ranked galaxies ($\Delta m_{12}$). Now we have an additional
dynamical evolution indicator, the fraction of IGL present in the
group.

In a first qualitative effort, we analyzed the relation between
these indicators in our sample. We restricted this analysis to the
groups studied here and in \citet{dar05} in order to have a
homogeneously analyzed sample and the analysis is only qualitative
due to small number of objects, only 6 groups.

\begin{figure}
\centering
\includegraphics[width=9.5cm]{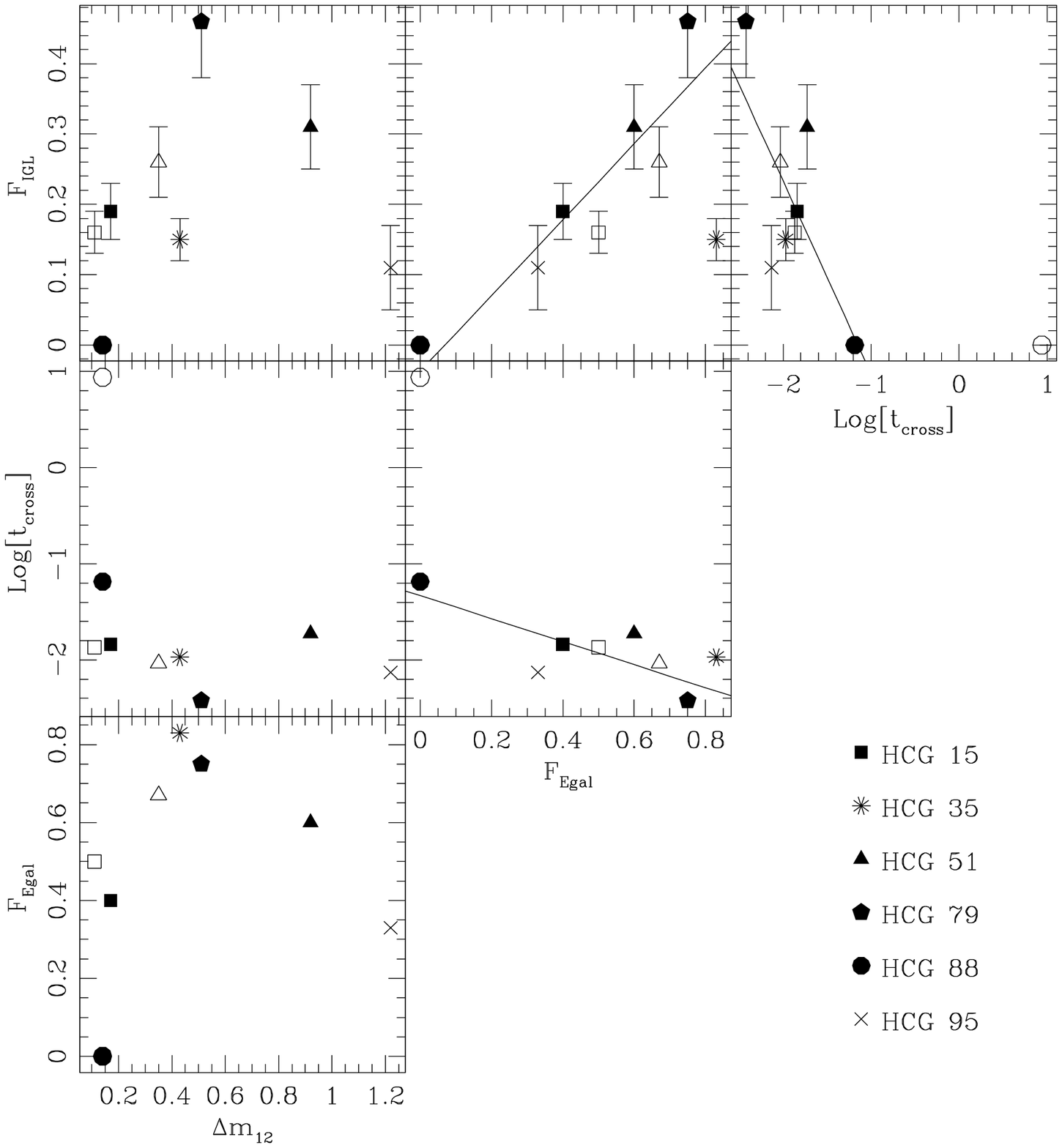}
\caption{Relations between different dynamical evolution indicators
($F_{IGL}$, $F_{Egal}$, $t_{cross}$ and $\Delta m_{12}$). The
continuous lines are linear fits to the points, excluding the ones
that clearly deviate from the main trend. The quintet configurations
for HCG 15 and HCG 51 are shown with filled symbols and the sextet
configurations are shown with the corresponding open symbol. The
``new'' crossing time value for HCG 88 is shown in filled symbol,
while the ``old'' value is shown in open symbol.
\label{figcorr}}
\end{figure}

We can see the relations between the fraction of IGL in the $B$
band ($F_{IGL}$), the fraction of early-type galaxies ($F_{Egal}$),
the crossing time ($t_{cross}$) and $\Delta m_{12}$, in
figure~\ref{figcorr}.  The fractions of IGL in the $B$ and $R$ band
are very similar, and for this kind of study do not present any
difference in the relation with the other indicators.

HCG 15 and 51 are presented with two data points each, for the
quintet and sextet configurations. For HCG 88 we have found new
recession velocity measurements \citep{nis00b}. The new measurements
increase the velocity dispersion of the group from $31$ to $93~{\rm
km~s^{-1}}$, allowing the estimate of the deprojected velocity
dispersion ($134~{\rm km~s^{-1}}$) and reducing the crossing time
from $8.7$ to $0.065~{\rm H_0^{-1}}$, therefore we included a ``new
point'' for this group in the relations between $t_{cross}$ and
other quantities.  This decrease in the crossing time do not alter
the conclusions about this object given in \citet{dar05}, since the
value is still considerably high.

The relations with $\Delta m_{12}$ show no clear tendency. Relations
between $F_{IGL}$, $F_{Egal}$ and $t_{cross}$ can be, in a first
approximation, described in a linear relation, if the clearly
discrepant objects are excluded. We excluded the ``old'' high
$t_{cross}$ value of HCG 88 from the fits for relations with this
quantity and HCG 35, that shows a very high fraction of early-type
galaxies, was excluded from the fits for relations with $F_{Egal}$.
These relations show the agreement, in a qualitative way, between
our new dynamical evolution indicator, the fraction of the total
light in the IGL and the fraction of early-type galaxies and the
group crossing time.

\section{Discussion and Conclusions} \label{discussion}

Before we start the discussion, we present a short summary of the
results of this work:

(1) We applied our wavelet analysis technique (the OV\_WAV package)
to study the IGL component in 3 further compact groups of galaxies,
completing a sample of 6 objects studied with this technique.

(2) HCG 15 had to be analyzed in two configurations (quintet and
sextet) and we were able to detect an IGL component that corresponds
to $19\pm4$\%/$16\pm3$\% of the total light in the $B$ band and
$21\pm4$\%/$18\pm4$\% in the $R$ band. The mean colour of the IGL
is $(B-R)_0 = 1.6\pm0.2$ consistent with the galaxies' mean colour.

(3) HCG 35 has IGL components which represent fractions of $15\pm3$\%
and $11\pm3$\% of the total light in the $B$ and $R$ bands,
respectively, with a mean colour $(B-R)_0 = 1.5\pm0.2$, bluer than
the mean colour of the galaxy component.

(4) HCG 51, which also had to be analyzed in two configurations
(quintet and sextet), presented $31\pm6$\%/$26\pm5$\% and
$28\pm5$\%/$24\pm5$\% of the total light in bands $B$ and $R$ in
the IGL component. The mean colour of the IGL component $(B-R)_0 =
1.5\pm0.2$ is compatible with the galaxy component mean colour.

(5) All the objects of this sample present irregular IGL components.
The analysis was made independently in the $B$ and $R$ bands and
the shape of the detected components agree in both bands, showing
the reliability of the method

(6) We find that the fraction of intragroup light, in the groups
studied so far, correlates, in a qualitative way, with other
indicators of dynamical evolution, such as the fraction of early-type
galaxies in the group and the crossing times.

As we already showed in \citet{dar05}, using the wavelet technique
OV\_WAV, we can detect the intragroup component present in compact
groups of galaxies in a very reliable way and without any ``a
priori'' information or assumptions about properties of the member
galaxies or of the image.

The components we detect correspond to about 40 to 75\% of the
luminosity of group first-ranked galaxies, which is equivalent to
15 to 35\% of a typical $M_{\star}$ luminosity \citep*[$M_B =
-21.9$][]{hun98}, where the first-ranked galaxies of the three
studied objects have magnitudes fainter than $M_{\star}$.  The
fraction of light in the IGL can be used as a measure of the amount
of interaction suffered by the member galaxies, which is expected
to be proportional to the number of crossings the galaxies suffered
since the group formation. By relating the amount of light in the
IGL with the expected amount of stripped matter in each crossing
we can estimate how many times the galaxy crossed the group. The
missing link here is the amount of matter stripped on average in
each crossing, which is sensitive to the distribution of the dark
matter and requires numerical simulations with very realistic initial
conditions to be determined.

The colours of the IGL are compatible with the colours of the galaxy
component.  In general this would indicate that the striping is due
to ongoing interactions, while redder colours for the IGL would
indicate an older stripping, since the stripped stars would evolve
passively and the galaxy would keep forming stars and maintaining
a bluer colour. In the case of compact groups, specially the ones
studied here, this relation can be misleading, since in these
early-type dominated groups, the galaxies do not typically have
star formation.

In the cases of HCG 35 and 51, the IGL is bluer than the mean colour
of the galaxy component, but still in agreement with each other.
The colours could indicate that some star formation regions were
or are being stripped from the galaxies, or that some star formation
could be occurring ``in situ''. As the groups have mostly early-type
galaxies, no star formation seems to be occurring there, even though
interactions are effectively happening, as can be noticed by the
presence of a red jet-like and probably also a very faint shell-like
structure in HCG 51. As in the case of HCG 79 \citep{dar05}, where
the colour difference was much larger than here, the cause for this
difference can also be the destruction of dwarf galaxies.

In the case of HCG 15, the IGL component is redder than the mean
colour of the galaxies' component, but also still in agreement, and
the fraction of early-type galaxies is lower. This could be the
case of stripped stars evolving passively, while the galaxies keep
forming stars. In this group also no signs of strong star formation
can be seen.

Ongoing interactions that do not form new stars, lead us to the
conclusion that the evolution of compact groups would be closely
related to the scenario of ``dry mergers''.  In the dry mergers
scenario \citep{dok05} most of the luminous field elliptical galaxies
are assembled not via high redshift gas rich disks, galaxy mergers
or monolithic collapse, but via merger of gas poor, bulge dominated
objects.

The shapes of the detected IGL components are very irregular (see
figures~\ref{figdifh15} to~\ref{figdifh51}) and cannot be described
with elliptical isophotes leading to large residuals. Also, the
fitted surface brightness profiles, lead to unphysical parameters
for the fit profiles (exponential, de Vaucouleurs, King-like core
model and Hubble-Reynolds), show that this component is irregular
in its whole extent.

The group's gravitational potential is dominated by the group's
dark matter halo and we assume that the light traces the mass.  If
the IGL is due to stripped stars that are distributed following the
group's dark matter dominated potential, the irregular shape of the
IGL lets us conclude that the groups are definitely not virialized
structures.  They would be in a compact configurations for long
enough to have part of the stars stripped from the galaxies and
dispersed in the group potential, but not enough to get in equilibrium.

The non-equilibrium situation, together with the small number of
galaxies, makes the estimate of group dynamical parameters, such
as mass and crossing time, very uncertain and sensitive to small
variations of the input data, as mentioned in section~\ref{grpparm}.

An important parameter is the total mass of the group, which can
only be determined dynamically, using the member galaxies, due to
the inexistence of other determinations, such as through X-ray
observations. In the case of compact groups of galaxies, the dynamical
mass estimators can only be used as an indicator, since the structure,
as we showed before, is not virialized. Also, they are only sensitive
to the mass inside the area occupied by the galaxies (the ``galaxies
circle''), which in the case of a compact group, can leave a
considerable part of the mass unaccounted for.

\citet*{mcc08} studied the properties of compact groups of galaxies
identified in a mock galaxy catalogue based on the Millennium Run
simulation \citep{spr05}.  These authors found that the member
galaxies are in the inner part of the common dark matter halo of
the group and that there is no correlation between the halo virial
radius and how the galaxies are concentrated towards the center of
the group.

We could approach this question in two extreme ways. One is assuming
that the dark matter halo dominates the mass in the whole extent
of the group and that the galaxies do not contribute significantly
to it. In this case the dark matter mass could be directly related
to the IGL luminosity, so that the fraction of mass unaccounted
would be proportional to the fraction of IGL outside the ``galaxies
circle''.  The missing fraction, in this case, would be about 60
to 70\% of the total mass. The second approach is to assume that
the galaxies dominate the mass in the inner part of the group and
that we have a constant $M/L$ in the whole structure. In this case
the missing fraction would not be larger than 10\% of the total
mass.

We are not able to pin down a number for the unaccounted mass, as
we are not able to pin down a number for the total mass itself using
dynamical estimators. The non-equilibrium state of the compact
groups, together with the fraction of the total mass that can be
measured by dynamical mass indicators, prevent us from a reliable
mass determination.  Other mass estimators, as X-rays, assume that
the structure is in equilibrium, what can lead to the same kind of
uncertainties.  For lensing estimates the small distance of the
groups and their low masses turn the detection of the shear signal,
if possible, very complicated. A recent work on the Coma cluster
\citep{kub07}, which is about the same distance as most of the HCGs,
determined its mass using weak lensing detected with deep, large
field of view, images from the SDSS. Even though the distances are
similar, the difference in total mass between Coma and the HCGs,
would still be a problem for detecting the shear signal. According
to \citet{men94b}, the HCGs would need a distance 10 times larger
to produce a detectable lensing signal.

In a qualitative way (our sample is too small for a more quantitative
result), we could show that the fraction of IGL agrees with other
dynamical evolution indicators such as the fraction of early-type
galaxies and the crossing time. The crossing time, as the mass, is
sensitive to small variations on the input parameters, as we showed
for the case of the previously studied group HCG 88. We can also
see that the magnitude difference between the first and the
second-ranked galaxies does not clearly correlate with any of the
other indicators.  \citet{hic92} shows that the crossing time
correlates with the fraction of spirals (complementary to the
fraction of early-type galaxies) and $\Delta m_{12}$ (figures 5 and
6 in their paper), but for objects in bins of crossing time. The
correlation using individual objects was not shown. So the correlation
of $\Delta m_{12}$, on average, could eventually be seen when we
have a considerably larger sample.

The new discordant velocity of HCG 15C, which has a much larger
radial velocity (by $2720~{\rm km~s^{-1}}$) than the median velocity
of the group, is a point to be considered with care.  Galaxies in
HCGs are bright enough for a good recession velocity determination
and both sources, the UZC and \citet{hic92}, have reliable redshift
determinations.

Altering the original group configuration, as excluding HCG 15C
(due to its new published recession velocity) or including HCG 51C
(extending the definition of ``discordant velocity'') puts a challenge
on defining which is the ``correct'' configuration for each of the
groups. In figure~\ref{figcorr} we can see the location of each of
the two configurations for both systems. For HCG 15, the relation
between $F_{Egal}$ and $t_{cross}$ leaves all the points too close
to the line for any consideration, as so does the relation between
$t_{cross}$ and $F_{IGL}$, but in the relation between $F_{Egal}$
and $F_{IGL}$, the quintet configuration shows a better agreement
with the main trend of the dynamical evolution indicators.  In the
case of HCG 51, the relation between $F_{Egal}$ and $t_{cross}$ and
between $t_{cross}$ and $F_{IGL}$ would favor the sextet configuration,
while the relation between $F_{Egal}$ and $F_{IGL}$, would favor
the quintet one.

Of course this kind of ``selection'' is very speculative, since
both configurations mostly agree between the errors. In a possible
more quantitative analysis, we can be able to extrapolated this
kind of relation as a test to verify if a galaxy that has an accordant
redshift, but do not seem to be involved in the interactions occurring
in the group, is a member of the group in a dynamical sense.

We can update our evolutionary sequence envisioned in \citet{dar05},
where we placed HCG 79 in the advanced stage of dynamical evolution,
HCG 95 in the intermediary stage and HCG 88 in the initial stage.
The three groups presented here would be placed in the intermediary
stage. All the objects have fraction of IGL between 10 and 30\%
(with the exception of HCG 51, when the quintet configuration is
considered it has 31\% of IGL fraction), the crossing times are in
the ``middle region'' (not as long as HCG 88 and not as short as
HCG 79) and the fraction of early-type galaxies is around 0.5 (except
for HCG 35 that has 0.83, clearly disagreeing with the other
indicators for this same group).

The fraction of early-type galaxies of HCG 35 is in clear disagreement
with the other indicators of dynamical evolution. This group also
shows an overabundance of H{\sc i}, when compared to the expected
H{\sc i} content of the group (estimated by the morphological type
and luminosity of its member galaxies). Its H{\sc i} content is
about 29\% higher than the expected, while HCG 15 and 51, in the
configurations published in \citet{hic92}, are H{\sc i} deficient
(about 70\% of the expected H{\sc i} is missing). However, the H{\sc
i} content of this group is dominated by the one Sb member galaxy
(all the other 5 members are early-type galaxies, where almost no
H{\sc i} is expected) and the H{\sc i} content of an Sb galaxy can
vary by a factor of 3, higher or lower than the typical value
\citep{rob94}. This shows that this group needs some extra attention
and studies.

The detection and analysis of IGL component in compact groups is
feasible and efficient, the results can be used as a dynamical
evolution indicator and can provide initial conditions to numerical
simulations in the search for a better understanding of group
formation and evolution and also to the formation and evolution of
clusters of galaxies.

\bigskip

Based on observations collected at the Centro Astron\'omico Hispano
Alem\'an (CAHA) at Calar Alto, operated jointly by the Max-Planck
Institut f\"ur Astronomie and the Instituto de Astrof\'{\i}sica de
Andaluc\'{\i}a (CSIC).  We would like to thank Carlos Raba\c ca and
Daniel Epit\'acio Pereira for the software development and support
and Hugo Capelato for discussions and support to this project.  We
thank the support from the Deu\-tsches Zen\-trum f\"ur Luft- und
Raum\-fahrt, DLR (project number 50 OR 0602), Deut\-sche
For\-schungs\-ge\-mein\-schaft, DFG (project number ZI 663/8--1)
within the Priority Program 1177 (SPP/DFG), Volkswagen Foundation
(project number I/76 520) and Funda\c c\~ao de Amparo a Pesquisa
do Estado de S\~ao Paulo, FAPESP (Project number 02/06881--4).  CMdO
would like to acknowledge support from the Brazilian agencies FAPESP
(projeto tem\'atico 01/07342-7), CNPq and CAPES. We would like to
thank the anonymous referee for the useful comments that considerably
improved this manuscript.

\bibliographystyle{mn2e} 

\bibliography{darocha}

\label{lastpage}

\end{document}